\documentclass[a4paper,11pt]{article}

\usepackage{amsmath}
\usepackage{authblk}
\usepackage{bbold}
\usepackage{cite}
\usepackage{fullpage}
\usepackage{graphicx}
\usepackage[utf8]{inputenc}
\usepackage{subcaption}
\usepackage[dvipsnames]{xcolor}
\usepackage{hyperref}

\title{The Effective Potential in Fermi Gauges Beyond the Standard Model}

\author[1,2]{Jonathan Zuk}
\author[1]{Csaba Bal\'azs}
\author[3]{Andreas Papaefstathiou}
\author[2]{Graham White}

\affil[1]{{\small School of Physics and Astronomy, Monash University, Melbourne 3800 VIC, Australia}}
\affil[2]{{\small Kavli IPMU (WPI), UTIAS, The University of Tokyo, Kashiwa, Chiba 277-8583, Japan}}
\affil[3]{{\small Department of Physics, Kennesaw State University, Kennesaw, GA 30144, USA}}

\date{\today}

\begin{document}

\maketitle
\begin{abstract}
 We derive the field-dependent masses in Fermi gauges for arbitrary scalar extensions of the Standard Model. These masses can be used to construct the effective potential for various models of new physics.  We release a flexible \texttt{Mathematica} notebook (\texttt{VefFermi}) which performs these calculations and renders large-scale phenomenological studies of various models possible.  Motivated by the debate on the importance of gauge dependence, we show that, even in relatively simple models, there exist points where the global minimum is discontinuous in the gauge parameter.  Such points require some care in discovering, indicating that a gauge-dependent treatment might still give reasonable results when examining the global features of a model.
\end{abstract}

\tableofcontents

\section{Introduction}

The observation of gravitational waves \cite{LIGOScientific:2016aoc, LISACosmologyWorkingGroup:2022jok, Caldwell:2022qsj} from the very early Universe will unlock a floodgate of information, providing us with an unprecedented richness of direct experimental probes for fundamental physics. Precision cosmology relying on gravitational waves is currently under development, and part of this effort involves the calculation of gravitational wave spectra from cosmological phase transitions. The typical first step of these calculations is the computation of the effective potential, the fundamental quantity that describes the relevant scalar sector. Consequently, precision gravitational wave cosmology for phase transitions begins with a precision calculation of the effective action \cite{Caprini:2019egz}.

Scalar fields play a vital role in this connection between cosmology and fundamental physics. The 2012 discovery of a Standard Model-like Higgs boson at the Large Hadron Collider (LHC)~\cite{ATLAS:2012yve,CMS:2012qbp} spectacularly confirmed the mechanism of electroweak symmetry breaking. The LHC, however, neither fully mapped out the Higgs potential, nor confirmed or ruled out the possible existence of additional scalars that may play a role in the electroweak phase transition. This is of importance, because the precise knowledge of the Higgs potential, or possibly the potential of a more extended scalar sector, is vital to understand the cosmological consequences of electroweak symmetry breaking. Beyond the possibility of gravitational waves, among these consequences are the stability of the electroweak vacuum \cite{krive1976vacuum, sher1989electroweak, espinosa1995improved, casas1995improved} and electroweak baryogenesis \cite{Morrissey:2012db, RevModPhys.71.1463, Cline:2006ts}. The precision cosmology of these phenomena demands the precision knowledge of the Higgs potential.

Furthermore, to understand these cosmological phenomena, it is imperative to not only be able to measure, but also to calculate the Higgs potential. The Higgs potential, of course, depends on the model Nature may have chosen beyond the Standard Model of elementary particles. Since this model is presently not known, the Higgs potential has been analysed at a substantial depth in the context of various new physics models.  A small selection of such models (accompanied by an incomplete selection of references) are: the Standard Model extended by a real scalar gauge singlet~\cite{OConnell:2006rsp, Profumo:2007wc, Barger:2007im, Espinosa:2011ax, Pruna:2013bma, Chen:2014ask, Kotwal:2016tex, Robens:2016xkb, Englert:2020gcp, Adhikari:2020vqo, Papaefstathiou:2020iag, Papaefstathiou:2021glr}, a complex singlet~\cite{Barger:2008jx,Coimbra:2013qq,Costa:2014qga,Jiang:2015cwa,Chiang:2017nmu,Dawson:2017jja,Cheng:2018ajh,Adhikari:2022yaa}, two real singlet~\cite{Robens:2019kga, Papaefstathiou:2020lyp,Robens:2022nnw}, a doublet~\cite{Gunion:2002zf,Davidson:2005cw,Branco:2011iw,Haller:2018nnx}, a singlet and a triplet~\cite{Bell:2020hnr}, a doublet and a real singlet \cite{Drozd:2014yla,Altenkamp:2018hrq}, two doublets~\cite{Keus:2013hya, Ivanov:2014doa,Maniatis:2014oza}, or a doublet and a triplet~\cite{Athron:2019teq}.

The classical Higgs potential receives substantial quantum corrections and it is essential to include these corrections in any reliable calculation.  
However, even with the advances in loop techniques, fixed-order one-, or two-loop, or resummed perturbative corrections, can be tedious to calculate depending on the model at hand. The situation worsens for the calculation of the effective scalar potential in the cosmological context. Here, perturbative calculations have to be performed in a thermal bath, and in the context of finite-temperature field theory, become even more demanding.  

The calculation of the effective potential is a subject of active research. Different authors have proposed various methods in the literature, with improvements as well as trade-offs, that aim to make these calculations more precise and/or more manageable. This has also prompted analyses of the uncertainties of the different parts of these calculations~\cite{Chiang:2018gsn,Guo:2021qcq,Athron:2022jyi}. The community appears to be divided regarding the choice of renormalisation scheme, resummation scheme, renormalisation scale, or gauge. The relative importance of including fixed-order or resummed perturbative corrections, or implementing gauge independence is also subject of discussion.

Gauge dependence becomes a particularly thorny issue in effective field theories. Here, unlike in perturbative calculations without a background field, gauge dependence is present from the outset, namely in the effective potential itself. As we show in this work, this gauge dependence can lead to qualitative differences in predictions. While gauge-independent calculations of the effective potential are being developed, they have not reached the maturity that would allow them to be employed to assess broad features of models, a task that would require a wide sampling of the model's parameter space. Thus, for the time being, it is important to assess the effect of gauge dependence of the effective potential in gauge-dependent methods. If one uses a gauge-dependent effective potential (motivated perhaps by convenience, or concerns about resummation), one should at the very least test the numerical sensitivity of observables to the gauge parameter. Doing so requires a consistent approach to evaluating the artificial gauge dependence -- that is, not including the zero-temperature vacuum expectation value in the gauge-fixing Lagrangian, but rather, using Fermi gauges.

The issue of gauge dependence of the effective potential was championed in ref.~\cite{Patel:2011th}, where a technique we refer to as the ``PRM method'' was developed for the analysis of finite temperature potentials.  It was embraced, improved and examined in detail in later papers, such as refs.~\cite{Chiang:2017nmu, Athron:2022jyi, Hirvonen:2021zej, Lofgren:2021ogg, Schicho:2022wty}.  Alternatively, in discussing gauge dependence and the method developed in ref.~\cite{Patel:2011th} the author of ref.~\cite{Kozaczuk:2015owa} expresses a philosophy, stating ``Although morally satisfying, the gauge-invariant approach has the disadvantage of sometimes neglecting numerically important contributions to the effective potential.''. Similarly, ref. \cite{Curtin:2016urg} suggests an improvement on resummation methods, but chooses to neglect gauge dependence in favour of focusing on resummation improvements while referencing (and discussing) the PRM method. Reference~\cite{Croon:2020cgk} (which shares a co-author with this paper) also gives similar reasons for comparing a gauge-dependent method with dimensional reduction, rather than PRM, while refs.~\cite{Papaefstathiou:2021glr, Papaefstathiou:2020iag} find gauge dependence to be sub-dominant. 
Finally, ref.~\cite{Garny:2012cg} has a philosophy similar to our work in varying the gauge parameter over a range to probe the numerical sensitivity.  
This is also done in ref.~\cite{Athron:2022jyi}, which concludes that gauge dependence is moderate through most of the parameter space of the $\mathbb{Z}_2$-symmetric scalar singlet extension of the Standard Model.

To improve gauge-dependent calculations of the effective potential, and to be able to better assess this gauge dependence, in this work we present a generic calculation of the field-dependent masses in the Fermi gauge. These masses are key ingredients for building the effective potential of a specific model. Although our caclulations will be at zero temperature, it is trivial to extend our analysis to the finite temperature case, as the derivation of these masses is the only non-trivial step. In an associated \texttt{Mathematica} notebook, we code the calculation of field-dependent masses, in the context of an arbitrary scalar extension of the Standard Model. The notebook also features the generic expression of the zero-temperature effective potential at tree level and at one loop (for a selected set of models), and finite-temperature correction terms.

We demonstrate the use of our generic calculation by applying it to two example cases: the Standard Model extended by a real scalar singlet, and by an additional scalar doublet. Intriguingly, we find pathological points in these two parameter spaces, where a small change (of 3) in the gauge parameter changes the location of the global minimum of the potential, rendering electroweak symmetry breaking itself gauge dependent.  

The remainder of this paper is structured as follows: in Section~\ref{sec:approach} we describe our approach for calculating the effective potential. We demonstrate this for two simple extensions of the Standard Model -- the Standard Model augmented by a real Scalar Singlet (SM+SS); and the Two-Higgs Doublet Model (2HDM) -- in Section~\ref{sec:models}. In Section~\ref{sec:num} we provide a benchmark for each of these models, where the global minimum changes discontinuously with the gauge parameter.  We discuss the implications and outlook in Section~\ref{sec:disc}. Basic information about our \texttt{Mathematica} notebook is provided in Appendix~\ref{app:notebook}.

\section{Effective potential in Fermi gauges} \label{sec:approach}

Whilst it is certainly the case that any observable quantity should be gauge independent, some useful quantities, such as the effective potential away from its tree-level minima, may be gauge dependent. In particular, the ratio of the gauge-dependent critical vacuum expectation value (vev) to the critical temperature is a frequently-used heuristic for the strength of the phase transition, correlating well with the sphaleron energy and the latent heat, while being more convenient to calculate. Furthermore, some observables calculated in a convenient manner from gauge-dependent quantities may themselves turn out to be gauge dependent.

Despite the theoretical distaste of a gauge-dependent observable, or even a gauge-dependent heuristic quantity, such a result may not be completely inadmissible, as long as the effect of gauge dependence is numerically small. This is particularly the case if one is primarily interested in performing rapid scans of a large region of parameter space. Incorporating resummation into a gauge-independent calculation requires at least some two-loop calculations\footnote{Recent work has shown how to construct a gauge-independent calculation that includes resummation in a more economical manner, see ref.~\cite{Schicho:2022wty}.} which might be excessive for a parameter-space scan. However, there remains the possibility of gauge dependence making a qualitative difference to the phenomenology. To assist the field in ascertaining the gauge dependence, both qualitative and quantitative, of a given model, we release a code that generates the effective potential in the Fermi gauge for an arbitrary model. We employ this code in what follows to establish the fact that there do indeed exist some parameter points that are qualitatively gauge dependent -- that is, there are dramatic discrete changes in the phenomenological predictions with a modest change in the gauge parameter.  We then use this to track the effects of gauge dependence and determine whether qualitative differences may arise, focusing initially on finding which is the deepest of the different minima.

The relevant gauge-fixing terms added to the Lagrangian for each gauge boson $A_i$ are of the form
\begin{equation}
    \mathcal{L}_{\rm{gf}} = -\frac{1}{2\xi_i}(\partial_\mu A_{i\mu}^a)^2.
    \label{eq:gaugefixing}
\end{equation}
We use the Fermi gauges rather than the generalised $R_\xi$ gauges sometimes employed for this purpose, since the latter method utilises different gauges for each value of the scalar field, the validity of which has been questioned~\cite{Arnold1992,LAINE1994173}. The generalised $R_\xi$ gauges are typically chosen for their cancellation of the off-diagonal Goldstone-longitudinal gauge boson terms, particularly as this results in much simpler propagators. However, in the Fermi gauges the effective potential may still be calculated with little difficulty, without requiring this cancellation which, moreover, becomes less of a concern once we are relying upon computational calculations.

We calculate the effective potential to 1-loop order using the background field method of ref.\cite{Jackiw1974}. Specifically, given a theory with a set of fields $\phi_i$ and action
\begin{equation}
    S[\phi]=\int d^4x~\mathcal{L}(\phi_a(x))\;,
    \label{eq:actiondef}
\end{equation}
the one-loop corrections to the effective potential are given by
\begin{equation}
    V_1(\hat{\phi})=-\frac{i}{2}\int\frac{d^4x}{(2\pi)^4}\ln\det i\mathcal{D}_{ij}^{-1}[\hat{\phi};k]\;,
    \label{eq:effpotcorrint}
\end{equation}
where $\hat\phi$ is a constant background and the inverse propagator may be evaluated using
\begin{equation}
    i\mathcal{D}_{ij}^{-1}[\hat{\phi};x,y] = \left.\frac{\delta^2 S[\phi]}{\delta\phi_i(x)\delta\phi_j(y)}\right|_{\phi=\hat\phi}\;,
\end{equation}
and then performing a Fourier transform. Taking the determinant is sufficient to readily identify the mass eigenvalues. In the $\overline{\rm{MS}}$ renormalisation scheme we obtain
\begin{equation}
    V_{1}(\phi)=\sum_{i}n_i\frac{m_i^4(\phi)}{64\pi^2}\left(\ln\left(\frac{m_i^2(\phi)}{\mu^2}\right)-k_i\right)\;,
    \label{eq:general1loopMSbar}
\end{equation}
where the sum runs over all fields in the theory, $n_i$ is the relevant multiplicity factor for each particle, which is taken to be negative for fermions, $m_i$ are the field-dependent masses, $\mu$ the renormalisation scale, and $k_i$ is given by
\begin{equation}
    k_i=
    \begin{cases}
        \frac{5}{6}, & \rm{gauge\;bosons} \\
        \frac{3}{2}, & \rm{otherwise}.
    \end{cases}
\end{equation}
This approach is demonstrated with concrete examples in the following section.

\section{Examples of application to specific models}\label{sec:models}

In this section we give an explicit calculation of two simple extensions of the Standard Model, with the more complex models reserved for the accompanying \texttt{Mathematica} code. Specifically, we work with the extension of the Standard Model by a real scalar singlet, and by a doublet scalar field (two-Higgs doublet model).

\subsection{The Standard Model plus a Real Scalar Singlet}

The addition of a real scalar singlet is the simplest extension to the scalar sector of the Standard Model (SM+SS). The SM+SS is also the simplest model where it becomes possible for the deepest minimum to qualitatively vary with the gauge. The derivation for the effective potential in this model in Fermi gauges was previously performed in~\cite{Papaefstathiou:2020iag}, and we review the calculation for completeness.

The most general renormalisable scalar potential in this model is
\begin{equation}
    V(H,S) = m^2 (H^\dagger H) + \frac{\lambda}{2}(H^\dagger H)^2 + K_1(H^\dagger H)S + K_2(H^\dagger H)S^2 + \frac{1}{2}m_s^2 S^2 + \frac{\kappa}{3}S^3 + \frac{\lambda_s}{2}S^4\;.
    \label{eq:SMSSpot}
\end{equation}
We decompose the Higgs doublet as
\begin{equation}
    H=\frac{1}{\sqrt{2}}\begin{pmatrix}\phi_1+i\phi_3\\
    \phi_2+v+i\phi_4\end{pmatrix}\;,
    \label{eq:HiggsDoubDecomp}
\end{equation}
where $v$ is a background field, and similarly expand the singlet around a background field $x$, $S=s+x$.  Collecting all the dynamical fields into a single vector as,
\begin{equation}
    \Phi = \begin{pmatrix}
    \phi_1 & \phi_2 & \phi_3 & \phi_4 & s & W_\mu^1 & W_\mu^2 & W_\mu^3 & B_\mu
    \end{pmatrix}^T\;,
    \label{eq:SMSSfields}
\end{equation}
the terms quadratic in the dynamical fields may be written as
\begin{equation}
    \mathcal{L} \supset -\frac{1}{2} \Phi^\dagger\Sigma\Phi + \mathcal{L}_{\rm{fermion}}\;,
    \label{eq:LagQuads}
\end{equation}
where $\Sigma$ is the inverse propagator matrix. Here, we may decompose $\Sigma$ as
\begin{equation}
    \Sigma = \begin{bmatrix}D^{ab} & M_\mu^a \\ M_\mu^{a\dagger} & \Delta_{\mu\nu}\end{bmatrix}
    \label{eq:Sigmadef}\;,
\end{equation}
where the scalar terms are given by 
\begin{equation}
    D^{ab}=\begin{bmatrix}
    -p^2 + d_H & 0 & 0 & 0 & 0 \\
    0 & -p^2 + d_H + \lambda v^2 & 0 & 0 & k_1v + k_2vx \\
    0 & 0 & -p^2 +d_H & 0 & 0 \\
    0 & 0 & 0 & -p^2 + d_H & 0 \\
    0 & k_1v + k_2vx & 0 & 0 & -p^2 + d_S
    \end{bmatrix}\;,
\end{equation}
the mixing scalar-gauge terms by 
\begin{equation}
    M_\mu^a =\begin{bmatrix}
    0 & \frac{i}{2}g_2vp_\mu & 0 & 0 \\
    0 & 0 & 0 & 0 \\
    \frac{i}{2}g_2vp_\mu & 0 & 0 & 0 \\
    0 & 0 & -\frac{i}{2}g_2vp_\mu & \frac{i}{2}g_1vp_\mu \\
    0 & 0 & 0 & 0
    \end{bmatrix},
\end{equation}
the gauge boson terms by
\begin{equation}
    \Delta_{\mu\nu}=\begin{bmatrix}
    \Delta_W-\frac{1}{4}g_2^2v^2g_{\mu\nu} & 0 & 0 & 0 \\
    0 & \Delta_W-\frac{1}{4}g_2^2v^2g_{\mu\nu} & 0 & 0 \\
    0 & 0 & \Delta_W-\frac{1}{4}g_2^2v^2g_{\mu\nu} & \frac{1}{4}g_1g_2v^2g_{\mu\nu} \\
    0 & 0 & \frac{1}{4}g_1g_2v^2g_{\mu\nu} & \Delta_B-\frac{1}{4}g_1^2v^2g_{\mu\nu}\\
    \end{bmatrix}\;,
    \label{eq:smssgaugegauge}
\end{equation}
and we have defined
\begin{align}
    d_H=&m^2 +k_1x + \frac{k_2}{2}x^2 + \frac{\lambda}{2}v^2\;, \\
    d_S=&m_s^2 + 2\kappa x + 6\lambda_sx^2 + \frac{k_2}{2}v^2\;, \\
    \Delta_W = &p^2g_{\mu\nu}-\left(1-\frac{1}{\xi_W}\right)p_\mu p_\nu\;, \\
    \Delta_B = &p^2g_{\mu\nu}-\left(1-\frac{1}{\xi_B}\right)p_\mu p_\nu\;.
\end{align}

Taking the determinant, we find that the gauge bosons and fermions\footnote{We consider only the top quark since it constitutes the dominant fermion contribution.} maintain the same masses as in the SM.  The Goldstone-like and physical Higgs scalar have masses given by
\begin{align}
    m^2_{1,\pm}=\frac{1}{2}\left(d_H\pm\sqrt{d_H(d_H-g_2^2\xi_W v^2)}\right)\;,\\
    m^2_{2,\pm}=\frac{1}{2}\left(d_H\pm\sqrt{d_H(d_H-(g_1^2\xi_B+g_2^2\xi_W) v^2)}\right)\;,\\
    m^2_{h,\pm}=\frac{1}{2}\left(d_H+d_S\pm\sqrt{(d_H-d_S)^2+4(k_1v + k_2vx)^2}\right)\;,
    \label{eq:SMSSmasses}
\end{align}
where the masses $m_{1,\pm}^2$ have a multiplicity of 2. Note that in this model the gauge dependence enters entirely through the masses of the Goldstone-like particles.

\subsection{The Two Higgs-Doublet Model}

As a further example, we consider the Two Higgs-Doublet Model (2HDM). To maintain some simplicity, we consider the case of only a softly-broken $\mathbb{Z}_2$ symmetry and no explicitly CP-violating terms. The relevant potential is
\begin{multline}
    V(H_1,H_2) = m_1^2(H_1^\dagger H_1) + m_2^2(H_2^\dagger H_2) - m_{12}^2(H_1^\dagger H_2 + h.c.) +\frac{\lambda_1}{2}(H_1^\dagger H_1)^2 + \frac{\lambda_2}{2}(H_2^\dagger H_2)^2 \\+ \lambda_3(H_1^\dagger H_1)(H_2^\dagger H_2) + \lambda_4(H_1^\dagger H_2)(H_2^\dagger H_1) + \frac{\lambda_5}{2}\left((H_1^\dagger H_2)^2 + h.c. \right)\;.
    \label{eq:2HMDpot}
\end{multline}
In general, the 2HDM admits the possibility of CP-violating and charge-breaking minima. However, since we are only interested in what happens with the deepest minimum, we restrict ourselves to vevs which are both CP-conserving and not charge-breaking, since if a vev of this nature exists, it is the global minimum. Proceeding as before, we expand the doublets as a set of scalar fields about these constant vacuum configurations as
\begin{equation}
    H_1=\frac{1}{\sqrt{2}}\begin{pmatrix}\phi_{11}+i\phi_{13}\\
    \phi_{12}+v_1+i\phi_{14}\end{pmatrix}\;, \quad
    H_2=\frac{1}{\sqrt{2}}\begin{pmatrix}\phi_{21}+i\phi_{23}\\
    \phi_{22}+v_2+i\phi_{24}\end{pmatrix}\;,
    \label{eq:2HiggsDoubDecomp}
\end{equation}
and create a vector of the fields
\setcounter{MaxMatrixCols}{12}
\begin{equation}
    \Phi = \begin{pmatrix}
    \phi_{11} & \phi_{12} & \phi_{13} & \phi_{14} & \phi_{21} & \phi_{22} & \phi_{23} & \phi_{24} & W_\mu^1 & W_\mu^2 & W_\mu^3 & B_\mu
    \end{pmatrix}^T\;.
    \label{eq:2HDMfields}
\end{equation}

We may again write the quadratic terms of the Lagrangian in the form of \eqref{eq:LagQuads} and \eqref{eq:Sigmadef}, where now the nonzero scalar terms $D^{ab}$ of the inverse propagator become
\begin{align}
    D^{11}=D^{33}=-p^2+m_{11}^2+\frac{1}{2}\left(\lambda_1 v_1^2 +\lambda_3 v_2^2\right)\;, \\
    D^{22}=-p^2+m_{11}^2+\frac{1}{2}\left(3\lambda_1 v_1^2 + \lambda_3 v_2^2 + \lambda_4 v_2^2+ \lambda_5 v_2^2\right)\;, \\
    D^{44}=-p^2+m_{11}^2+\frac{1}{2}\left(\lambda_1 v_1^2 + \lambda_3 v_2^2 + \lambda_4 v_2^2- \lambda_5 v_2^2\right)\;, \\
    D^{55}=D^{77}=-p^2+m_{22}^2+\frac{1}{2}\left(\lambda_2 v_2^2 +\lambda_3 v_1^2\right)\;, \\
    D^{66}=-p^2+m_{22}^2+\frac{1}{2}\left(3\lambda_2 v_2^2 + \lambda_3 v_1^2 + \lambda_4 v_1^2+ \lambda_5 v_1^2\right)\;, \\
    D^{88}=-p^2+m_{22}^2+\frac{1}{2}\left(\lambda_2 v_2^2 + \lambda_3 v_1^2 + \lambda_4 v_1^2- \lambda_5 v_1^2\right)\;, \\
    D^{15}=D^{51}=D^{37}=D^{73}=-m_{12}^2+\frac{1}{2}\left(\lambda_4v_1v_2+\lambda_5v_1v_2\right)\;, \\
    D^{26}=D^{62}=-m_{12}^2+\lambda_3v_1v_2+\lambda_4v_1v_2+\lambda_5v_1v_2\;, \\
    D^{48}=D^{84}=-m_{12}^2+\lambda_5v_1v_2\;,
\end{align}
the mixed terms are given by
\begin{equation}
    M_\mu^a =\begin{bmatrix}
    0 & \frac{i}{2}g_2v_1p_\mu & 0 & 0 \\
    0 & 0 & 0 & 0 \\
    \frac{i}{2}g_2v_1p_\mu & 0 & 0 & 0 \\
    0 & 0 & -\frac{i}{2}g_2v_1p_\mu & \frac{i}{2}g_1v_1p_\mu \\
    0 & \frac{i}{2}g_2v_2p_\mu & 0 & 0 \\
    0 & 0 & 0 & 0 \\
    \frac{i}{2}g_2v_2p_\mu & 0 & 0 & 0 \\
    0 & 0 & -\frac{i}{2}g_2v_2p_\mu & \frac{i}{2}g_1v_2p_\mu
    \end{bmatrix}\;,
\end{equation}
and the gauge terms retain the same form as \eqref{eq:smssgaugegauge} but with the substitution $v^2\to v_1^2+v_2^2$. The explicit form of the masses are in this case the roots to polynomials of up to quartic order which, since they cannot be expressed concisely, we omit here and leave to the notebook. In general, polynomials of arbitrarily-high order may be encountered, resulting from matrices with large dimensions. Finding the determinant analytically of these large matrices is nontrivial -- in this 2HDM case the matrix is naively $24\times24$ when expanding the Lorentz indices. However, some simplification results from the fact that the transverse and longitudinal components of the gauge bosons contribute in a separable way, which our notebook takes advantage of. Finally, note that, in this model we must also specify which of the doublets the fermions couple to.

\section{Numerical results and gauge dependence of the deepest minimum} \label{sec:num}

In this section we use the potentials derived in section~\ref{sec:models} to investigate qualitative changes that occur in the effective potential as a result of varying the gauge parameter. We find that it is possible for the global minimum of a potential to switch between two local minima with the gauge choice in these models, and present a benchmark where this occurs for each model.  Since there are some issues with the convergence of perturbation theory for an arbitrarily large gauge parameter \cite{LAINE1994173,Garny:2012cg}, we restrict our analysis to consider values of the gauge parameter up to $\xi \leq 3$.  This leads to a relatively small effect compared to the barrier height, see sections \ref{sec:smssnum},\ref{sec:2hdmnum}.  However, it is worth noting a much greater effect may be produced if allowing a larger change in the gauge parameter sometimes used in the literature, such as the ranges considered in refs. \cite{Patel:2011th,Athron:2022jyi}.

\subsection{The Standard Model plus a Real Scalar Singlet}\label{sec:smssnum}

\begin{figure}
    \centering
    \begin{subfigure}{0.48\textwidth}
    \includegraphics[width=0.95\linewidth]{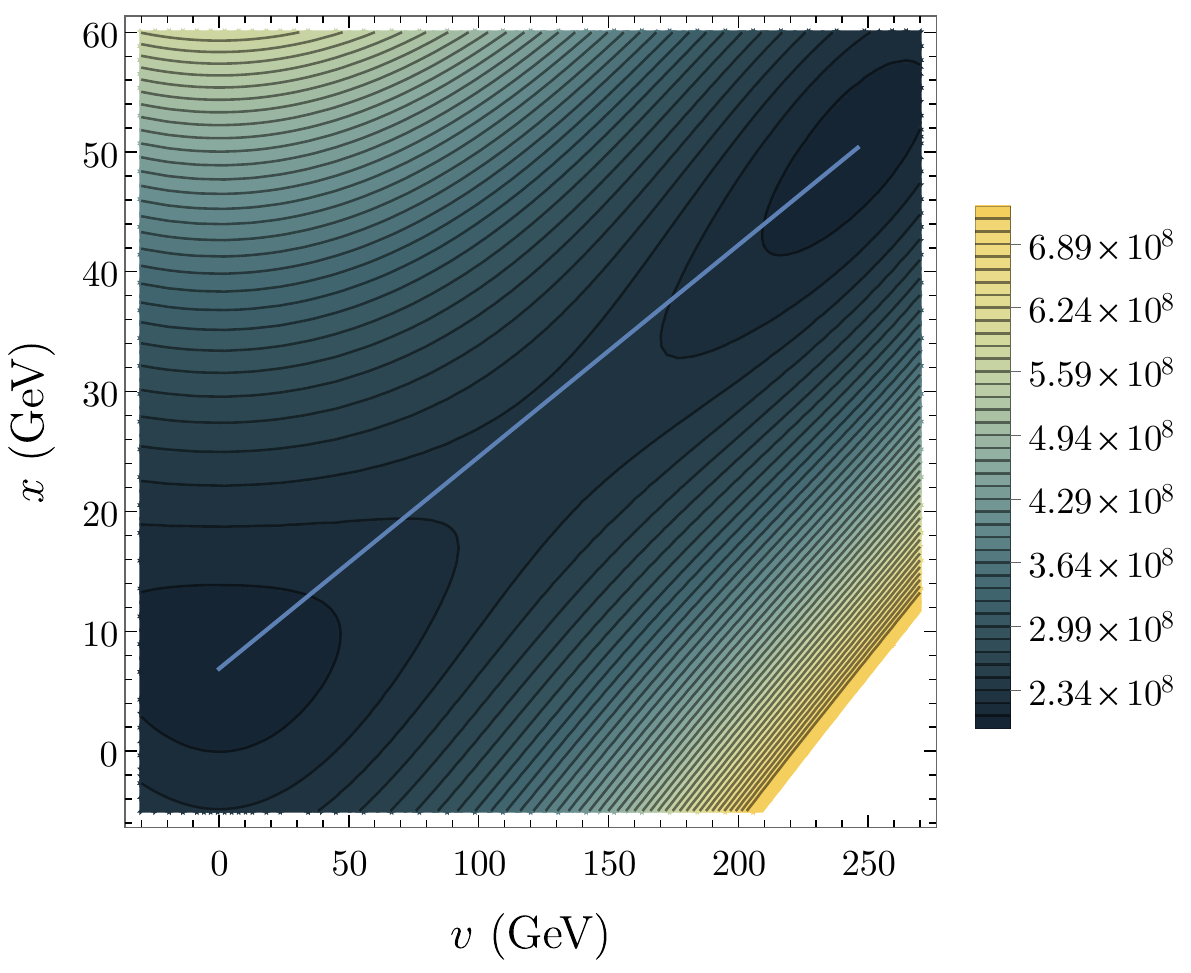}
    \end{subfigure}
    \begin{subfigure}{0.48\textwidth}
    \includegraphics[width=0.95\linewidth]{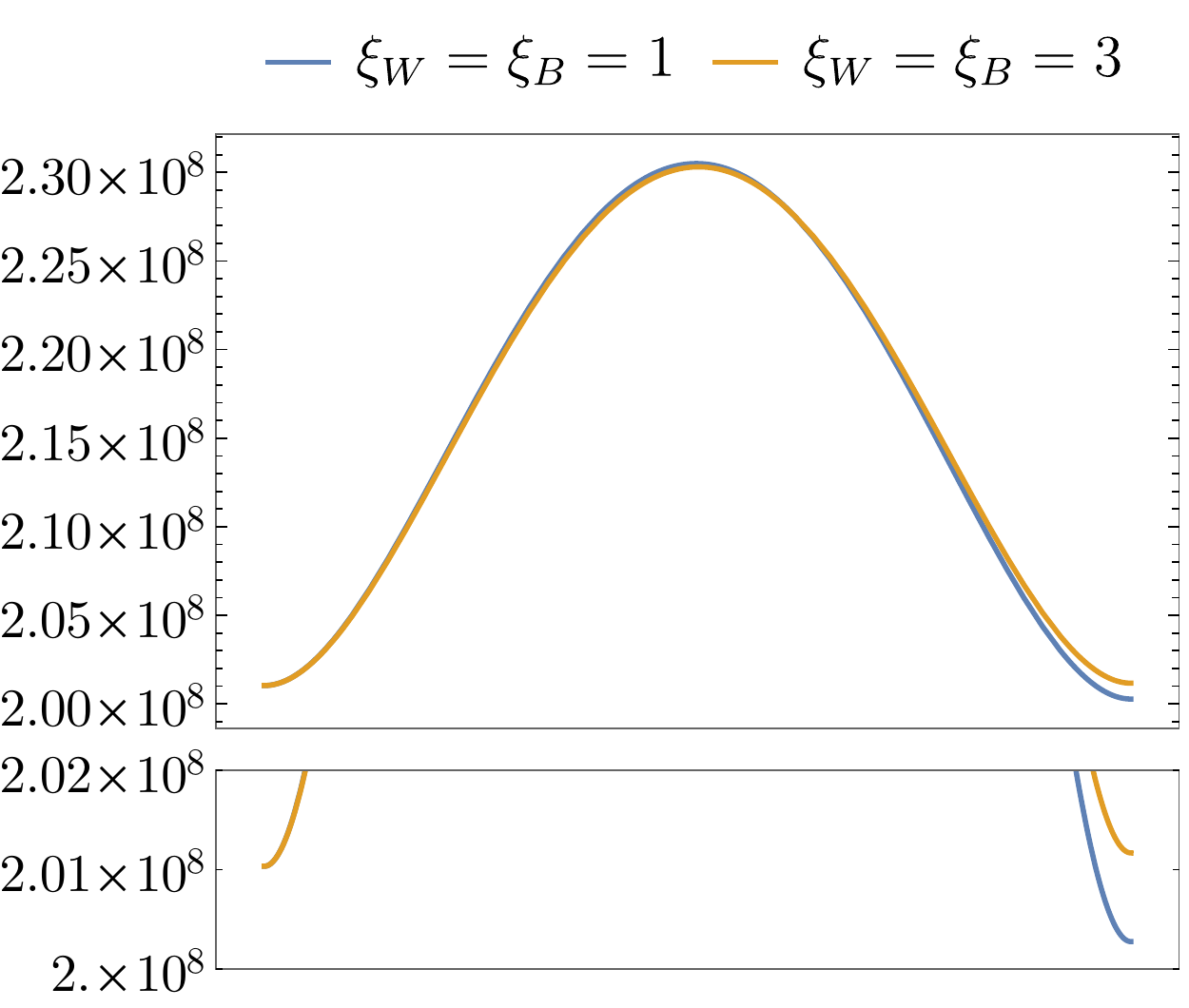}
    \end{subfigure}
    \caption{Left panel: Contour plot of the effective potential for the parameter values given in table \ref{tab:SMSSparamvals} and in Feynman gauge. The blue line connects the minima and shows the path of the blue curve of the right panel in field space. Right panel: the effect of the gauge in determining the minimum demonstrating the value of the potential along the line connecting the minima.  Note that since the position of the minima also changes with the gauge parameter, the orange curve takes a slightly different path in field space to that shown with the blue line in the left panel.}
    \label{fig:smsspot}
\end{figure}

\begin{table}[]
    \centering
    \begin{tabular}{c|c}
        $m^2$ & $-6500~\rm{GeV}^2$ \\
        $\lambda$ & 0.0576 \\
        $K_1$ & $-620~\rm{GeV}$\\
        $K_2$ & 6.6 \\
        $m_s^2$ & $260000~\rm{GeV}^2$\\
        $\kappa$ & $-480$ GeV\\
        $\lambda_s$ & 1 \\
        $y_t$ & 1 \\
        $g_1$ & 0.357 \\
        $g_2$ & 0.652
    \end{tabular}
    \caption{Parameter values for a benchmark in the SM+SS model where the deeper of two minima changes with changing the gauge parameters by 3. The parameters reproduce the observed values for the mass of the Standard Model Higgs boson and vev of 125 GeV and 246 GeV respectively \cite{ParticleDataGroup:2022pth}.}
    \label{tab:SMSSparamvals}
\end{table}
With the specific parameter values show in table \ref{tab:SMSSparamvals}, it can be easily verified that, in Feynman gauge, this results in a global minimum at one loop with $v=246.01~\rm{GeV}$ and $x=50.4$~GeV and the lighter of the physical Higgs fields having a mass $m_h=124.99$~GeV in this minimum.  That is, we match the first and second derivatives of the effective potential at one loop.  We also have the mass of the heavier Higgs state of 738.1 GeV and a mixing angle of 0.189.  Note, however, that there is another local minimum with $v=0$ GeV. The situation is shown in figure \ref{fig:smsspot}. The values of the potential at these two minima are nearly degenerate, with values of $2.00\times10^8 \ ({\rm GeV})^4$ and $2.01\times10^8 \ ({\rm GeV})^4$, respectively.

If we now change the values of the gauge parameters to $\xi_B=\xi_W=3$, we find that the minimum at 0 becomes the global minimum of the theory.  This is a consequence of a gauge-dependent contribution at one loop to the effective potential being generated by the Goldstone-like masses.  This means the electroweak minimum may be modified, while the symmetric minimum remains gauge independent, and since the minima are nearly degenerate, this can be enough to change the global minimum of the potential.

Note that this point did not require a large fine tuning, with the parameters needing to be specified to only two significant figures. The value of $\lambda$ is only specified to three digits to ensure that the Higgs mass constraint is met.  This indicates these points are at least common enough to be taken seriously, though a gauge-dependent scan may still be of utility for understanding macroscopic features of a potential.  

\subsection{The Two Higgs-Doublet Model}\label{sec:2hdmnum}

\begin{table}[]
    \centering
    \begin{tabular}{c|c}
        $m_1^2$ & $-4960~\rm{GeV}^2$\\
        $m_2^2$ & $-8700~\rm{GeV}^2$\\
        $m_{12}^2$ & 0 GeV\\
        $\lambda_1$ & 0.1\\
        $\lambda_2$ & 0.3\\
        $\lambda_3$ & 0.88\\
        $\lambda_4$ & $-0.1$\\
        $\lambda_5$ & $-0.05$\\
    \end{tabular}
    \caption{Parameter values for a benchmark in the 2HDM model where the deeper of two minima changes with changing the gauge parameters by 20.  The values of the top Yukawa, the gauge couplings, and the standard Higgs mass and vev constraints are the same as for Table \ref{tab:SMSSparamvals}.}
    \label{tab:2HDMparamvals}
\end{table}

We also find that a discrete change in the global minimum may occur when varying the gauge parameter in the 2HDM with significant qualitative differences, and present a benchmark where this occurs. For a concrete example, we further simplify to the case of $\mathbb{Z}_2$ symmetry. We also choose to couple the fermions to just the $H_2$ doublet to avoid the possibility of tree-level flavour-changing neutral currents in order to retain phenomenological relevance. We then consider the example parameter values of table~\ref{tab:2HDMparamvals}.

In the Landau gauge we again have a global minimum where only the second doublet obtains a vev, matching the SM phenomenology for the Higgs vev and the lighter Higgs mass. Another local minimum occurs only in the first doublet. However, when increasing the gauge parameters we find that this second minimum becomes the global minimum. In this case we would obtain massless fermions. Once again, we find that for this particular case we are unable to draw any conclusions as to whether this parameter point presents a phenomenologically-plausible candidate.

However, in this case the difference in the depths relative to the barrier height is smaller than in the SM+SS benchmark.  All the minima have depths of about $-1.24\times10^8$~GeV${^4}$ and when changing from Landau gauge to $\xi_W=\xi_B=3$, we go from one minimum being deeper by $1.3\times10^5$~GeV${^4}$ to the other by $1.6\times10^5$~GeV${^4}$. Whether this means only a small region of parameter space is affected or not requires a scan to determine with certainty, which is left to future work.
\begin{figure}
    \centering
    \begin{subfigure}{0.48\textwidth}
    \includegraphics[width=0.95\linewidth]{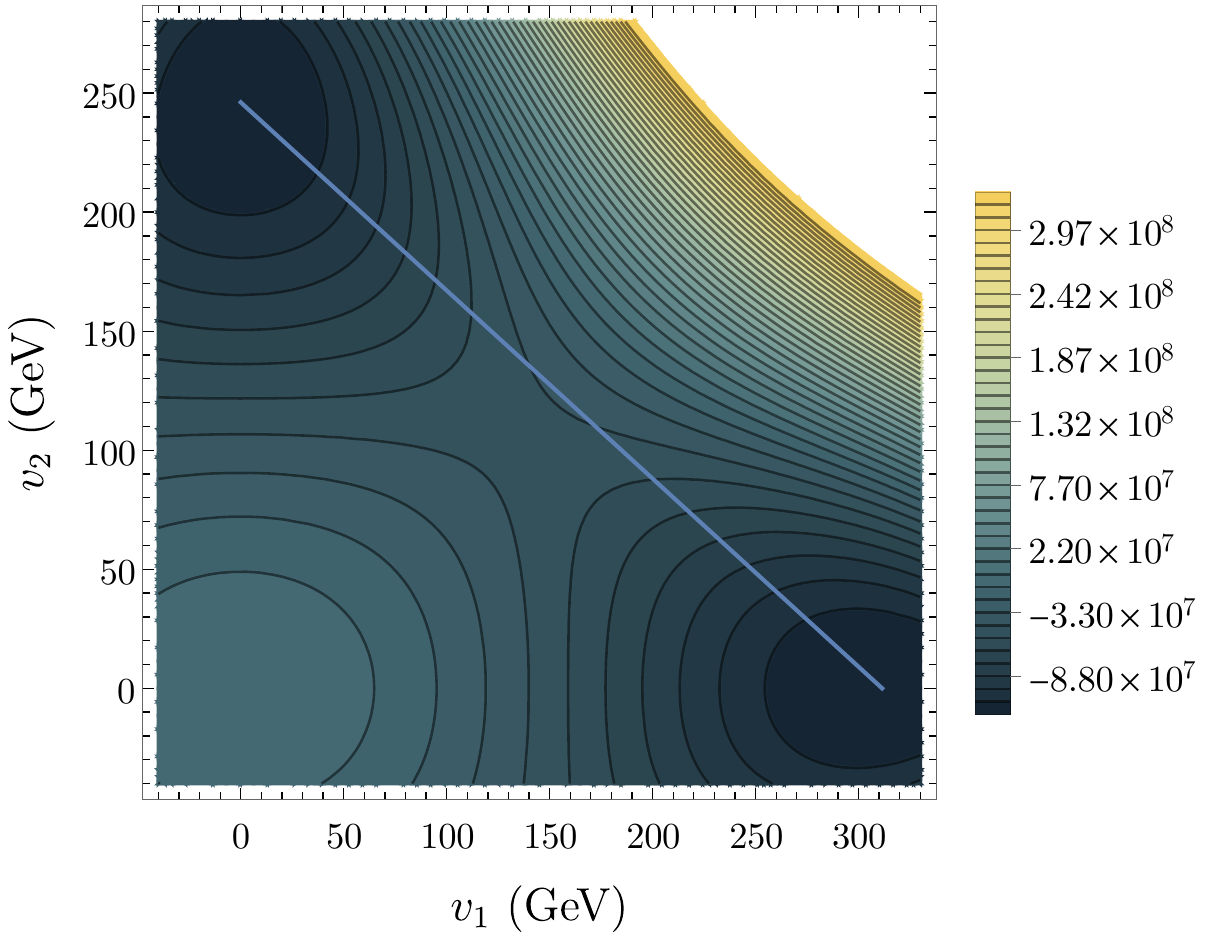}
    \end{subfigure}
    \begin{subfigure}{0.48\textwidth}
    \includegraphics[width=0.95\linewidth]{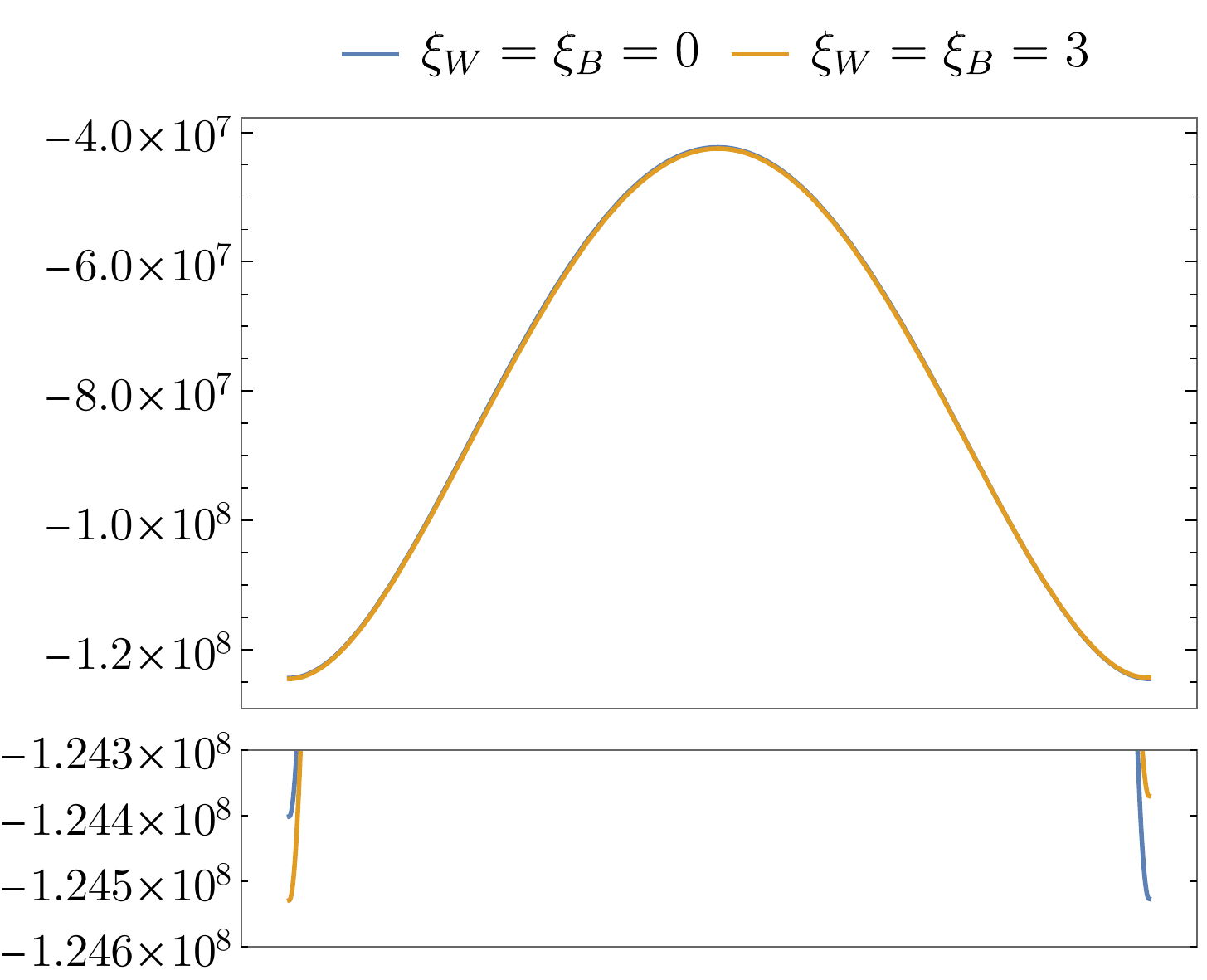}
    \end{subfigure}
    \caption{Left Panel: Potential for the benchmark values given in Table \ref{tab:2HDMparamvals} and in Landau gauge, with the blue line denoting the field direction connecting the two minima. Right panel: The potential along the lines connecting the minima for two different values of the gauge parameter as per figure \ref{fig:smsspot}.}
    \label{fig:2hdmpot}
\end{figure}

\section{Discussion} \label{sec:disc}

In this work we presented a calculation of the the zero temperature effective potential, up to one loop in Fermi gauges, in the context of an arbitrary scalar extension of the Standard Model.  We coded this calculation in a publicly available \texttt{Mathematica} notebook, which also includes finite temperature corrections.  Using this code we examined a few points in the parameter space of two models: the Standard Model extended with a gauge singlet and the two-Higgs doublet model.  

In both of these models, we have seen that points which appear to have correct phenomenology in one gauge may appear in another to be in fact unphysical, with very different qualitative behaviour. That we have found benchmarks where this is the case due to the global minimum changing with the gauge in two very simple extensions of the standard model, suggests that this is likely to be a general feature of many models.  This provides an additional reason for caution in using gauge-dependent effective potentials, though the fact that these points require some modest fine tuning suggest that a gauge-dependent scan could still have some utility for seeing the broad features of a whole model.

However, this being said, it remains unclear at this stage how large a region of parameter space is affected by this issue. Our search for such points suggests that these points may be relatively rare, although not in need of excessive fine tuning. Furthermore, from our benchmarks, it would appear that the amount of points so affected varies considerably with the model.  A statistically meaningful comment on how common these points are would require a full scan, model by model which we leave to future work.

While throughout the bulk of the parameter space gauge dependence may be small compared to uncertainties arising from renormalisation scheme and scale choice, or choice of resummation method \cite{Athron:2022jyi}, our benchmarks demonstrate that gauge dependence of the effective potential can be important for selected parameter points. Due to this, a calculation of the effective potential that completely ignores gauge dependence cannot be considered reliable, and in general it is desirable to check the severity of gauge dependence.  Using the Fermi gauges is particularly suited for this because it reliably captures the gauge dependence of the effective potential. Using our \texttt{Mathematica} package, it is possible to calculate the effective potential in the Fermi gauges, for a wide range of models, which enables users to quickly and efficiently perform a parameter scan within a given model. Such a scan can reliably reveal problematic regions of the parameter space where gauge dependence is important.

\section*{Acknowledgements}

This work was supported by the Australian Research Council Discovery Project grant DP210101636. AP acknowledges support by the National Science Foundation under Grant No. PHY 2210161. The work of GW is supported by World
Premier International Research Center Initiative (WPI),
MEXT, Japan. GW was supported by JSPS KAKENHI
Grant Number JP22K14033.

\appendix

\section{Mathematica notebook}\label{app:notebook}

\texttt{VefFermi} is a \texttt{Mathematica} notebook which allows for the calculation of the effective potential in the Fermi gauges for arbitrary scalar extensions of the Standard Model, with any number of additional fields, from the scalar potential and the background fields.  Numerical evaluation is fast and takes about 0.2 seconds for the the Coleman-Weinberg and thermal corrections, which are the most complicated functions.  It can be downloaded from \url{https://github.com/JonathanZuk/VefFermi}.

Optimal functionality requires \texttt{Mathematica 13.0} or above, however, workarounds exist which are compatible with at least version 12 and possibly earlier releases.

Current features include functions which calculate (in the Fermi gauges):
\begin{itemize}
    \item The inverse propagator: \texttt{massMatrix}, which takes the scalar potential and background fields as inputs
    \item The (squared) field dependent masses: \texttt{masses}, which takes the output of \texttt{massMatrix} and parameter values as input
    \item The tree level effective potential: \texttt{vTree}, which has the same inputs as \texttt{massMatrix}
    \item The Coleman-Weinberg (1-loop) corrections to the effective potential: \texttt{vColemanWeinberg}, which has the same inputs as \texttt{massMatrix} and \texttt{vTree}, with the addition of the (squared) fermion masses in terms of the background fields, and parameter values
    \item The thermal corrections to the effective potential of the form
    \begin{equation}
        V_1^\beta=\sum_{i\in\rm{bosons}}{n_i\frac{1}{2\pi^2\beta^4}J_B\left(m_i^2\beta^2\right)}-\sum_{i\in\rm{fermions}}n_i\frac{1}{2\pi^2\beta^4}J_F\left(m_i^2\beta^2\right),
        \label{eq:general1loopthermal}
    \end{equation}
    where $J_B$ and $J_F$ are the thermal bosonic and fermionic functions respectively: \texttt{vThermal}, which has the same inputs as \texttt{vColemanWeinberg} with the addition of temperature.
\end{itemize}

Further instructions can be found within the notebook.  This includes how to use the above function to calculate these quantities both numerically and analytically.  It also includes a number of models which are already implemented as examples.  These are the extensions of the Standard Model by: a real singlet; a doublet; a triplet; and two real singlets.

\bibliographystyle{unsrt}
\bibliography{main} 

\end{document}